\documentclass[9pt,twocolumn,twoside]{pnas-INA} 

\templatetype{pnasresearcharticle}

\usepackage{amsthm,amssymb,amsmath}
\usepackage{tikz-network}
\usepackage{sansmath}
\usepackage{xspace}
\usepackage{lipsum}
\usepackage{xcolor}

\usepackage{coffee4}

\graphicspath{{../figures/}}

\theoremstyle{definition}

\definecolor{gray}{RGB}{66,66,66}
\definecolor{blue}{RGB}{27,72,105}
\definecolor{orange}{RGB}{255,139,0}

\renewcommand{\eqref}[1]{Eq.~\ref{eq:#1}\xspace}

\title{Predicting kills in Game of Thrones using network properties}

\author[a, b]{Jaka Stavanja}
\author[a, b]{Matej Klemen}
\author[a, c]{Lovro Šubelj} 

\affil[a]{University of Ljubljana, Faculty of Computer and Information Science, Ljubljana, Slovenia}
\affil[b]{\{js8927|mk3141\}@student.uni-lj.si}
\affil[c]{lovro.subelj@fri.uni-lj.si}

\begin{abstract}
TV series such as HBO's Game of Thrones have seen a high number of dedicated followers, mostly due to the dramatic murders of the most important characters.
In our work, we try to predict killer and victim pairs using data about previous kills and additional metadata. 
We construct a network where two character nodes are linked if one killed the other and use a link prediction framework to evaluate different techniques for kill predictions. 
Lastly, we compute various network properties on a social network of characters and use them as features in conjunction with classic data mining techniques.
Due to the small size of the dataset and the somewhat random kill distribution, we cannot predict much with standard indices alone, although using them in conjunction with additional rules based on degrees works surprisingly well.
The features we compute on the social network help the classic machine learning approaches, but do not yield very accurate predictions. 
The best results overall are achieved using indices that use simple degree information, the best of which gives us the Area Under the ROC Curve of $0.875$.
\end{abstract}

\dates{This manuscript was compiled on \today}
\doi{\url{arxiv.org}}

\begin{document}
\maketitle
\thispagestyle{firststyle}
\ifthenelse{\boolean{shortarticle}}{\ifthenelse{\boolean{singlecolumn}}{\abscontentformatted}{\abscontent}}{}

\section{\label{sec:introduction}Introduction}
\begin{figure}[] \centering
	\includegraphics[width=0.5\textwidth]{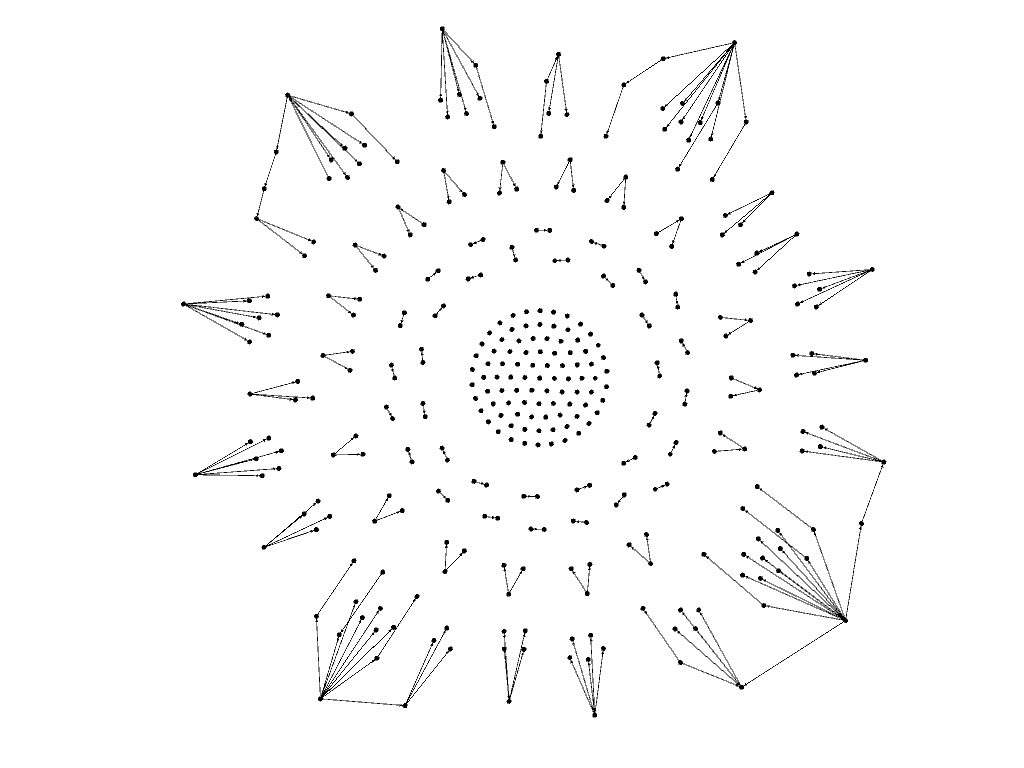}
	\caption{{\bf Game of Thrones kills network.}}
	\label{fig:got_network}
\end{figure}

With the ever increasing popularity of the TV series Game of Thrones, coming mostly from it's incredible plot twists and deaths of main characters, the question arises whether we can predict those deaths from a network analysis point of view. 
If we are able to predict the deaths from the data, collected from previous episodes, that means that the author is very predictable, which might not be the best thing in terms of the show being entertaining. 
To predict the deaths, we construct a network illustrated in Figure~\ref{fig:got_network}, where nodes are characters in the show (along with other entities that are able to kill another character, such as a horse or a dragon) and connect two if one has murdered the other.
We then try to predict whether a certain link between two nodes in the network happened (removing the link from the network beforehand).
Because we are dealing with temporal data, we also remove links that appear in the network after the link currently being predicted. 
We use different approaches to assign scores to the pairs, for example  different link prediction indices such as the preferential attachment index~\cite{LNK07} or the Adamic-Adar index ~\cite{AA03}. 
We also construct additional features based on various network properties and additional metadata and see how this influences the results. 
The network properties for the features used in machine learning approaches are then computed on a different social network of the Game of Thrones characters, where nodes representing characters are connected if the characters appear somewhat close in the original books' story.

The rest of the paper is structured as follows. 
In Section \ref{sec:related_work} we provide an overview of existing literature on our studied topic. 
In Section \ref{sec:methods} we describe the methods that we use in our work. 
In Section \ref{sec:results} we provide the results of our work, which we then discuss in Section \ref{sec:discussion}.
In Section \ref{sec:conclusion} we summarize the work that was done and provide some possible future improvements.

\section{Related work}
\label{sec:related_work}

We use link prediction as a way to infer new links between nodes in a graph using different network properties.
The field of link prediction research started evolving around trying to either generate synthetic networks or infer missing data for network-like data structures. 
Later on, that evolved into more of a recommendation type of approach (for example, trying to recommend friendships on social networks), based on the same principles as we use to calculate probabilities of new links.
One of the most important articles on the network properties that we can use to infer new links is written by Barab{\'a}si \& Albert and features exploring the phenomenon of preferential attachment~\cite{BA99}. 
The work on this property and other founding principles for link prediction is then briefly described inside the article by the same authors~\cite{BA02}. 
The next explored thing is missing data~\cite{K06} along with further studies on missing links and spurious networks~\cite{GS09}.
An extensive survey on link prediction is written by L{\"u} \& Zhou~\cite{LZ11} where the whole field along with all the better-known indices to date is presented and the authors show how the classic problems in link prediction didn't use enough network properties or community structure, which was explored by Girvan \& Newman~\cite{GN02}. We use some of their findings in our research.
One of the most well known indices to calculate the likelihood of new links appearing in a network is the common neighbors index~\cite{N01}~\cite{K06}, implying that the more common neighbors nodes have, the more likely they are to form a link. 
Some other popular choices include the cosine distance index (also named the Salton index)~\cite{MS83}, the Jaccard index~\cite{J01}, the preferential attachment index~\cite{BA99} and the Adamic-Adar index~\cite{AA03}. 
For our experiments, we pick some of the most used indices.


For Game of Thrones specifically, some research has already been done in terms of finding which aspects of the show resonate with the viewer count the most and how real the characters' interactions are done by modeling the show's houses with a network and exploring structural balance~\cite{LA17}. 
Further studies determined who has the best strategic position in the show's world~\cite{BS16}, but the only article touching on death prediction studies~\cite{ABSSAR2018} for the show used Cox's proportional hazard model~\cite{C72}, which didn't explore any network structure properties, but rather used a regression approach to determine which factors introduce a higher risk of mortality through time. But the killers were not taken into consideration here, we just know how likely a character is to die, thus we cannot really compare this approach with ours.
This gives us a unique opportunity to try and use network properties along with metadata to better predict kills in the mythical world of the series, but instead of only predicting how likely someone is to die, we can also predict both the killers and victims, on which there has not yet been much research.

\section{\label{sec:methods}Methods and network properties}

In this section, we describe the framework used to test our link prediction methods. 
We also provide descriptions of some classic link prediction methods based on well-known properties of networks. 
Then, we introduce a new social network, described in Subsection~\ref{sec:feature_extraction}, which connects characters if they appear close in the original story from the books. 
We compute various network properties from that network and use an automatic node embedding technique to compute features for use in traditional machine learning approach for classification.

\subsection{Link prediction using the kills network}
\begin{figure*}[] \centering
	\includegraphics[width=0.7\textwidth]{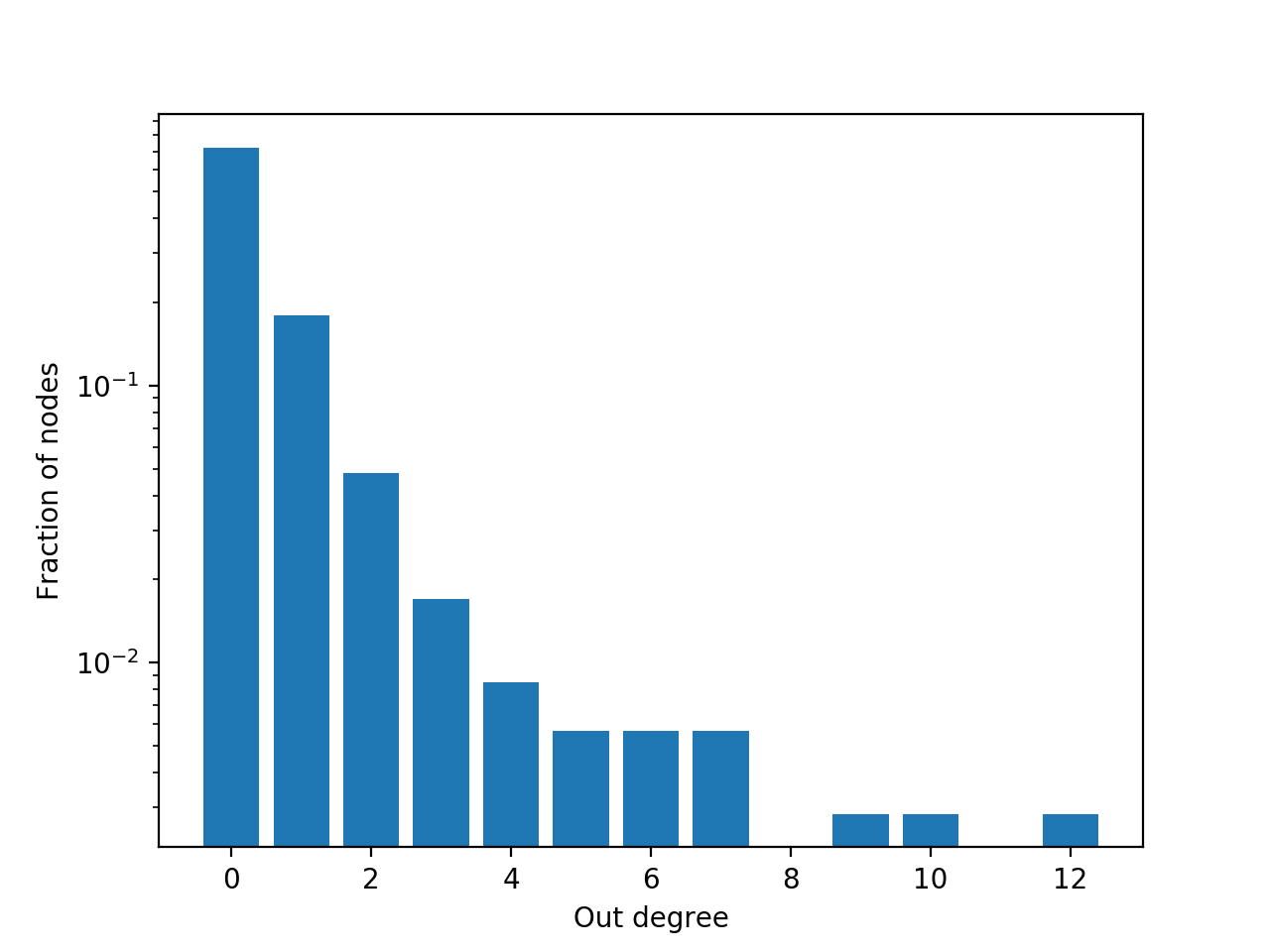}
	\caption{{\bf Out-degree distribution for the kills network.}}
	\label{fig:out_degree_distribution}
\end{figure*}

The network of kills is constructed of directed links between pairs of nodes $i$ and $j$, where node $i$ represents a character that killed the character represented by node $j$. 
The network is very small and also very sparse. 
It has $353$ nodes and $194$ links. 
By looking at the visualization of the network in Figure~\ref{fig:got_network}, we can see that most kills appear outside somewhat bigger connected components and do not seem to be attributed to hubs (i.e. high degree nodes).
We can still observe a few hubs and connected components around them, where a lot of kills seem to occur, but in the majority of cases there are just two nodes involved in the kill, which on their own form a small two-node connected component.
That gives us a clue as to which methods might predict kills better than others. 
We can construct an index that predicts links based on nodes' out degrees. 
Since the in-degrees in the networks could be either zero or one (meaning alive or dead) we cannot get any additional information from that besides whether a person can still kill someone or not (when they are already dead). 
The out-degree of a node can potentially be of use, since people that kill a lot of people might also tend to kill more people, and those who never killed anyone might not be inclined to murder or kill.
A plot of the out-degree distribution (using a logarithmic scale for the fractions of nodes) is shown in Figure~\ref{fig:out_degree_distribution}.

In this subsection, we propose an evaluation framework and different indices that we use for the link prediction.

\subsubsection{\label{sec:framework}Evaluation framework for link prediction}
To test how well our link prediction techniques work on our network, we construct a framework and provide a brief description of it here.
It takes our prediction index function and the network as the input and outputs the Area Under the ROC Curve (AUC) value.
The prediction index function assigns a score $s_{ij}$ to every link between two nodes $i$ and $j$ that is being tested. 
A high score implies a high likelihood that the link exists in the network and a low score implies a small likelihood of the link's existence.

The core of the testing framework is the logic, which removes links from the network and then tries to predict how likely the links we have removed are to form as the network evolves through time. 
To use as much information as we can, we choose an episode and remove links from that episode (e.g. episode $30$) and onwards from the graph. 
Then, we predict the links (i.e. kills) at that time using the information about kills from the previous episodes. We can then clone the original graph, remove links from the next episode in the chronology (i.e. episode $31$) and predict links for that episode using the information from all the episodes before (including information from episode $30$ in this example). By predicting links in this way, we are not using the data from the future to predict past links and we are using a lot more information than if we were to remove links from a certain episode onwards and just try to predict links for multiple episodes in one single iteration.

The idea of the framework's core implementation is as follows:

\begin{enumerate}
    \item Iterate through every episode in the range from $30$ to $60$, and at each step do:
    \begin{enumerate}
        \item $L_{P} \gets $ remove node links $\left(i, j\right)$ $\in L$ after the current episode.
        \item Compute $s_{ij}$ for $\left(i, j\right)$ $\in L_{P}$ at the current episode time.
        \item Save the scores to $S_P$.
    \end{enumerate}
    \item $L_{N} \gets $ randomly sample $|L_P|$ unlinked nodes $\left(i, j\right)$ $\notin L$.
    \item Compute $s_{ij}$ for $\left(i, j\right)$ $\in L_{N}$ and save scores to $S_N$. 
    \item Compute AUC by comparing the scores assigned to positive samples $S_P$ and scores assigned to negative samples $S_N$. When comparing scores, we assign to $b$ the amount of times when the score from $S_P$ is bigger than the one from $S_N$, and assign to $e$ the number of times when the two scores are equal. This is counted across a random sample of $|L_P|$ pairs. Then, we compute the AUC as $\frac{b+e/2}{|L_{P}|}$.
\end{enumerate}

\subsubsection{Alive index}
We create a type of a baseline index by looking at the network's high level properties. 
We check if the killer has an in-degree of zero and the target has an in-degree of zero (i.e. killer is alive and target has not been killed yet).
If the endpoints of a link satisfy these conditions, the value of the index is $1$, otherwise it is $0$.

\subsubsection{\label{sec:preferential_attachment_index}Preferential attachment index}
Real world networks tend to have a scale-free degree distribution due to a phenomenon known as preferential attachment~\cite{BA99}. 
The preferential attachment index~\cite{LNK07} is defined as $s_{ij} = k_ik_j$, where $k_i$ is the degree of node $i$. For our problem, we use out-degrees only, as only they provide useful information.
The idea behind the index is in the preferential attachment process --- nodes are more likely to connect with nodes that have a high degree, thus a link between two nodes with high degrees should be assigned a high score $s_{ij}$. 
We modify this definition slightly and define a modified preferential attachment index as $s_{ij} = k_i^{(out)}k_j^{(out)}$, where $k_i^{(out)}$ is the out-degree of node $i$.
The logic behind that is that the more kills one has to their record, the more they are inclined to murder and vice-versa.
Additionally, we include the in-degree information in a modified version of the preferential attachment index: if the source node or the target node has an in-degree larger than zero (either the killer or the victim is already dead), a very negative score ($-\infty$) is returned.
Otherwise, the regular version of the index gets computed.

\subsubsection{\label{sec:adamic_adar_index}Adamic-Adar similarity index}
In real world networks, links tend to appear between nodes that have a lot of common neighbors~\cite{WS98}. 
But due to preferential attachment, nodes tend to connect to higher degree nodes more likely, thus making that neighbor less useful for predicting new links between two nodes. 
The Adamic-Adar similarity index~\cite{AA03} takes the high degree neighbors into account and is defined as $s_{ij} = \sum_{x \in \Gamma_{i} \cap \Gamma_{j}} \frac{1} {\log k_x}$, where $\Gamma_{i}$ is the neighborhood of node $i$ and $k_{x}$ is the degree of node $x$.
Similarly, as for the preferential attachment index, we also include a modified version of the Adamic-Adar index, where we take into consideration whether the killer or the victim is already dead.

\subsubsection{\label{sec:community_index}Community index}
New links in e.g. social networks tend to appear between members that are inside of the same community and only rarely between members of different communities~\cite{GN02}. 
We can find densely linked communities where people are killing each other the most and treat the new links in those communities as more likely to occur than the ones outside communities. 
So that we do not ignore links between communities, we can also count the amount of links that occur between two communities and model the inter-community densities. Let \{$C$\} be the set of communities output by the Leiden modularity optimization algorithm~\cite{TWVE18}, and $c_{i}$ the community of node $i$. Then,
\[
s_{ij} = \left\{\begin{array}{lr}
    m_i/{n_i \choose 2}, & \text{when } \ c_{i} = c_{j}\\
    m_{ij} / (n_i \cdot n_j), & \text{when } \ c_i \neq c_j
    \end{array}\right\},
\]
where $n_i$ is the number of nodes in the community of node $i$, $m_i$ the number of links within community of node $i$ and $m_{ij}$ is the number of links between communities of nodes $i$ and $j$.
As is the case for previous two indices, we include a modified version of the community index as well, returning a very negative score when the killer or victim are already dead.

\subsection{\label{sec:feature_extraction}Feature extraction for machine learning approaches}
\begin{figure*}[] \centering
	\includegraphics[width=0.7\textwidth]{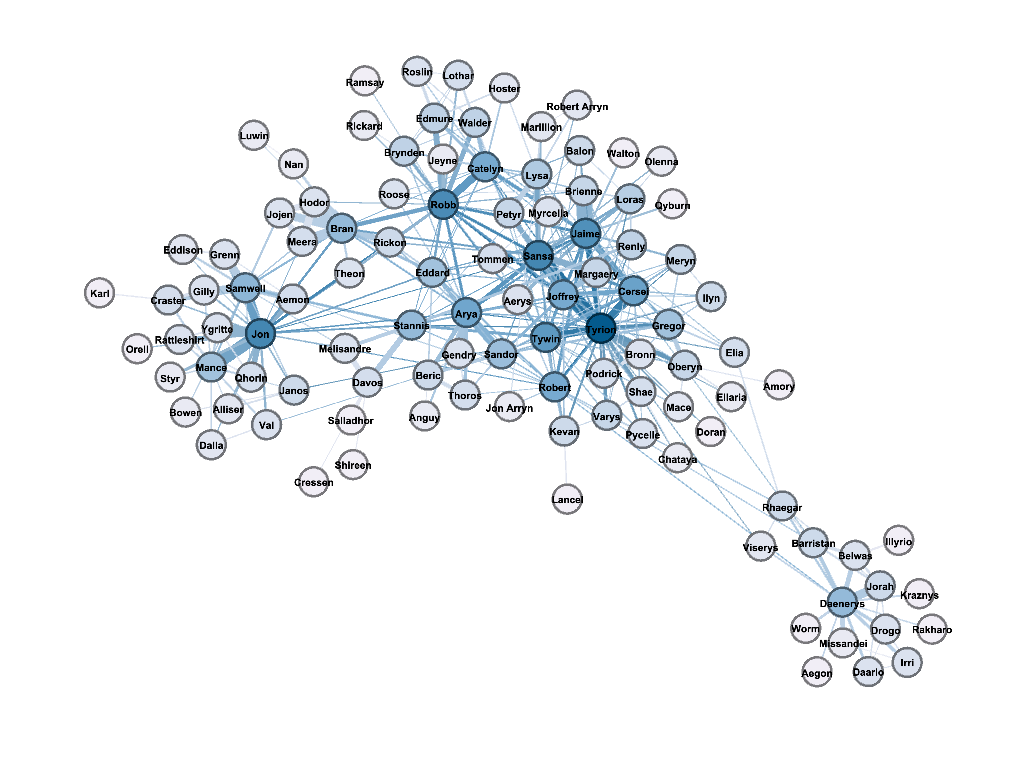}
	\caption{{\bf Game of Thrones social network.}}
	\label{fig:got_social_network}
\end{figure*}

Besides using the classic index-based link prediction techniques, we also make use of a machine learning-based approach, in which we construct features from network properties and additional metadata, and use them to train a classifier.
The data used for training the classifier includes properties of the kills network such as PageRank scores for all characters and the basic kills from the original dataset.
Computing a score like PageRank~\cite{BP98} on the kills network would be pointless as the nodes have an in-degree of at most one, so the scores would be high for those who already died. 
We can take a different network of characters into account for this particular case. 
An online repository of sample networks, found at \url{https://github.com/melaniewalsh/sample-social-network-datasets}, contains a sample of the Game of Thrones social network, which creates an edge between two character nodes if they appear within a $15$-word distance in the original books. 
Since the shows are vastly influenced by the books (at least very strongly up to episode $60$ to which our kills data is collected) we can use properties from that network as well to gather some additional information.
This network has $107$ nodes and $352$ edges, and is shown in Figure~\ref{fig:got_social_network}.

\subsubsection{Standard machine learning framework}

The framework we use for predicting links using machine learning is similar to the framework that is described in Section \ref{sec:framework}.
We split the procedure for getting scores for test links into two parts --- first we obtain the scores for positive examples and then the negative examples.

To obtain the scores for positive examples, we first choose some episode, whose kills we are currently trying to predict.
These links make up our current test set.
Links that appear in the episodes after the selected one are ignored, as we must not predict the past based on future events.
The kills that happened prior to the selected episode make up our current training set.
In addition to that, we sample the same amount of negative examples in order to make the training set balanced.
The classifier is then trained on the training set and used to predict scores for chosen test examples.
This is repeated for multiple chosen episodes and at the end we obtain scores for $|P|$ positive examples, where $P$ contains all the kills that have happened in the chosen episodes.

For negative examples, we take all the kills from our dataset and sample the same amount of non-kills to form a training set.
The test set consists of $|P|$ randomly sampled non-kills from the entire dataset.
These test examples are sampled in a way that they do not overlap with negative examples in the training set.

For obtaining the scores, we use three different models: K-nearest neighbors (KNN), logistic regression and support vector machine (SVM)~\cite{HTF09}.
The rest of the framework remains the same as the one described in Section \ref{sec:framework}. 

\subsubsection{PageRank}
The first feature we use is the PageRank score~\cite{BP98}. 
We can calculate it to find the most important characters as they will hopefully have the highest scores.
Being more important could mean two things --- either you are very important and thus have a lot of security by guards and other helpers, so you are very unlikely to die, or the exact opposite. The opposite would imply that since this series is oriented around taking the power from others, you are more likely to die if you are very important, which seems more plausible since the show thrives on the sudden deaths of the more popular characters.
A very low PageRank score of some character would then also mean that they are very unlikely to have their death portrayed, since the viewers and readers don't care or have forgotten about the least important people in the story.
We use two PageRank-based features, the first being \textit{killer\_pagerank} and the second \textit{victim\_pagerank}, since we are trying to predict killer and victim pairs.
The top 5 scoring characters after PageRank calculations are Tyrion ($0.055$), Jon ($0.045$), Daenerys ($0.041$), Jaime ($0.037$) and Sansa ($0.036$). From our knowledge of the show, we can safely claim that these are some of the most, if not the most important characters, so we can then be sure that these scores make sense in terms of importance. We assign a mean of all the PageRank scores for each character that is found in the kills dataset, but is not present in the social network.

\subsubsection{Betweenness centrality}
Another measure that we extract from the social network as a feature is the betweenness centrality~\cite{F77}. 
This score measures how many shortest paths from two different nodes go through a certain other node. 
That means that the higher the score, the more control over a big portion of the network a node has (i.e. is one of the nodes that can make the network split quickly). 
That should also yield a measure of importance --- if one character wants to reduce the power of a part of a network, they can disconnect it from the biggest component by eliminating nodes with high betweenness centrality. 
Again, we observe that the five top scoring people are also the most important characters in the story: Jon ($0.230$), Robert ($0.209$), Tyrion ($0.198$), Daenerys ($0.157$) and Robb ($0.127$). For every character from the kills dataset that is not in the social network, we use a mean of betweenness centrality scores, similarly to the PageRank scores.

\subsubsection{Community detection}
Simply using the house that a character belongs to as a feature could perhaps help.
The intuition behind this is that there might be more kills occuring between members of different houses than between members of the same house. 
Since our dataset has a lot of undefined values for the houses, we can find communities in the social network using an algorithm such as the Leiden modularity optimization~\cite{TWVE18}. 
We can assign a label of the community to each character and see if that helps us with the prediction by creating some sort of indicator between which groups the kills occur. 
After running the community detection algorithm on the social network, we can see that we obtain meaningful results (meaningful to people who watch the show) in terms of alliances.
The network gets partitioned into five big communities. 
For example, we can see that Khal Drogo, Daenerys Targaryen, Aegon Targaryen and Jorah Mormont all fall into the same community, even though they do not originate from the same houses, but as we know from watching the show, they are allies. Every character from the kills dataset that is not included in the social network gets assigned to a dummy community.

\subsubsection{\label{sec:feature_extraction}Automatic feature extraction for machine learning approaches}

In addition to handcrafting features we also try an approach using automated feature extraction, specifically node2vec~\cite{grover2016node2vec} to obtain node and link features.
We use these in place of previously handcrafted features.

For a given node, node2vec constructs a feature representation (embedding) that aims to preserve the network neighbourhood properties in a vector space of fixed dimensionality.
Depending on parameters $p$ (return parameter) and $q$ (in-out parameter), the algorithm performs different types of biased random walks in order to represent the node's neighbourhood.
Setting $p$ to a low value encourages a search that is local to the given node, while setting $q$ to a low value encourages a more explorative search.
The sampled neighbourhoods are then used to estimate a feature representation that maximizes the probability of observing these neighbourhoods for the given node.
We obtain a link embedding by concatenating together the embeddings of source and target node.
If a node has no embedding, a generic embedding for an unknown node is assigned to it.
We train the embeddings on the social network, capturing character co-ocurrences.
Because we are dealing with a very small dataset, we cannot afford to tune the hyperparameters reliably. 
Following the findings of authors of the method, we set the parameters to $p = 2$ and $q = 0.5$ since these settings are shown to bias the embeddings to capture homophily.
We fix the node embedding size to $16$.

\section{\label{sec:results}Results}

For our experiments we select and remove links that are associated with kills that happened in seasons four to six.
There are $114$ of those, to which we add $114$ randomly selected unlinked nodes and compute the AUC based on these examples.
We repeat this process five times to account for the randomness in selection of negative examples and provide the mean AUC and its standard deviation. 
We support the AUC scores with precision and recall scores as well, since AUC only measures how well a randomly selected positive example can be distinguished from a randomly selected negative example.
For the classic link prediction techniques, we classify examples with scores strictly higher than zero as positive and the others as negative. 
For the machine learning approach, every example with a score greater than or equal to $0.5$ is classified as positive and the others as negative.
We then calculate the precision as 
$$precision = \frac{\#true \ positives}{\#true \ positives + \#false \ positives}$$ 
and recall as 
$$recall = \frac{\#true \ positives}{\#true \ positives + \#false \ negatives}.$$
The obtained results are shown in Table~\ref{tab:preliminary-results}.

Results show that the community index achieves the best result in terms of AUC, $0.875$. The best precision and recall scores are obtained using the alive index, which are $0.822$ and $0.930$. Other indices using the death info achieve similar results, however their precision and recall scores are different.

The Adamic-Adar and community index achieve $0$ precision and recall because no links get classified as positive. For the community index, this is due to the network components being very disconnected, while for the Adamic-Adar this is due to the fact that nodes cannot have common neighbors as kills are only attributed to one person, meaning two killers cannot kill the same victim.

Other indices that don't use information about deaths achieve AUC scores around $0.5$. This implies that they perform no better than if links were classified randomly.

\begin{table}[]
\centering
\caption{Results for four link prediction methods: preferential attachment index, Adamic-Adar index, community index and alive index. The table shows the mean AUC, precision and recall and their standard deviation over five runs. The $\dagger$ symbol marks the versions of indices where a check is first performed if the killer or the victim is already dead.}
\label{tab:preliminary-results}
\begin{tabular}{@{}cccc@{}}
\toprule
Method    & AUC & Precision & Recall \\ \midrule
preferential attachment & 0.503 (0.020) & 0.522 (0.113) & 0.087 (0.000) \\
preferential attachment $\dagger$ & 0.862 (0.015) & 0.654 (0.123) & 0.070 (0.000) \\
Adamic-Adar & 0.500 (0.000) & 0.000 (0.000) & 0.000 (0.000) \\
Adamic-Adar $\dagger$ & 0.872 (0.028) & 0.000 (0.000) & 0.000 (0.000) \\
community index & 0.500 (0.000) & 0.000 (0.000) & 0.000 (0.000) \\ 
community index $\dagger$ & \textbf{0.875} (0.018) & 0.000 (0.000) & 0.000 (0.000) \\
alive index & 0.863 (0.032) & \textbf{0.822} (0.020) & \textbf{0.930} (0.000) \\ \bottomrule
\end{tabular}
\end{table}

The standard machine learning approaches came out to be a little bit better than random when using the base kills dataset using only the out-degrees of characters as a feature, with AUC scores ranging from $0.556$ to $0.650$ as shown in Table \ref{tab:ml_basic}.

\begin{table}[]
\centering
\caption{Results for three classic machine learning methods (using the basic dataset): K-nearest neighbors, logistic regression and support vector machine (SVM). The table shows the mean AUC, precision and recall and their standard deviation over five runs.}
\label{tab:ml_basic}
\begin{tabular}{@{}cccc@{}}
\toprule
Method                          & AUC                    & Precision              & Recall        \\ \midrule
KNN                             & \textbf{0.632} (0.048) & 0.641 (0.024)          & 0.453 (0.020) \\
logistic regression             & 0.596 (0.036)          & \textbf{0.797} (0.010) & 0.400 (0.013) \\
SVM                             & 0.556 (0.066)          & 0.688 (0.024)          & \textbf{0.523} (0.007) \\ \bottomrule
\end{tabular}
\end{table}

By adding network features (PageRank, betweenness and community identifier for each character) we improve the general performance of all the classifiers and obtain a better top score of $0.686$ by using SVM and the handcrafted features. 
The features created by node2vec yield worse AUC and precision scores and higher recall scores than the handcrafted features.
All the final results for the machine learning approaches are shown in Table \ref{tab:ml_augmented}.

\begin{table}[]
\centering
\caption{Results for three classic machine learning methods (using the augmented dataset): K-Nearest-Neighbors, Logistic regression and Support Vector Machine (SVM). The table shows the mean AUC, precision and recall and their standard deviation over five runs.}
\label{tab:ml_augmented}
\begin{tabular}{@{}cccc@{}}
\toprule
Method              & AUC           & Precision     & Recall        \\ \midrule
KNN                 & 0.659 (0.035) & 0.725 (0.016) & 0.453 (0.004) \\
Logistic Regression & 0.658 (0.033) & \textbf{0.816} (0.016) & 0.418 (0.013) \\
SVM                 & \textbf{0.686} (0.058) & 0.719 (0.058) & 0.456 (0.056) \\
KNN (node2vec)                  & 0.650 (0.044) & 0.595 (0.052)          & \textbf{0.702} (0.036) \\
logistic regression (node2vec)  & 0.640 (0.025)          & 0.600 (0.051)          & 0.684 (0.028) \\
SVM (node2vec)                  & 0.605 (0.091)          & 0.642 (0.062)          & 0.632 (0.022) \\
\bottomrule
\end{tabular}
\end{table}

\section{\label{sec:discussion}Discussion}
We see that the sparsity of the network and its size (less that $400$ nodes) make link prediction on such a small network very inaccurate in most cases.

Since the original network of kills does not seem to have any community-like structure, it is very hard to predict kills based on community-based link prediction methods, such as the community index. 
The modularity optimization algorithm finds more than $100$ communities with no links between them, which means that the modeling of densities between communities does not give us any information, since the probability of a link occurring between communities is zero.
That is a solid foundation for the claim that the authors have done a good job by \textit{not} creating a very obvious structure of the kills, where someone would kill a lot of people from e.g. their opposing house, making it easy to predict that there will be another similar kill occurring in the following episodes which the viewers have not yet seen. 

We cannot achieve good results by constructing indices that try to predict kills using out-degrees only, since the majority of nodes have a very low out-degree (as can be observed on Figure~\ref{fig:out_degree_distribution}).

When we add the death information, we observe that the modified Adamic-Adar index performs as good as the alive index, since it represents the same idea. We know that two nodes cannot have a common successor, since a character can only die once. They can only have one common predecessor, however that implies that both the killer and victim are already dead. By taking the death information into consideration, we automatically decide that when one character is already dead, there will be no link. That makes the common neighborhood factor in the index irrelevant, making it decide the outcome based only on whether a character is alive or not.

The indices that use the death information achieve the best results because of the way the network is constructed.
Nodes either have an in-degree of zero or one (depending on whether they were already killed or not).
When sampling positive examples in our framework, we remove the edge for that positive example, decreasing the in-degree of the target node from one to zero.
When sampling negative examples, we do not remove any edges.
However, because our original network is constructed from deaths in the series, most of the characters in our network have died at some point. 
Therefore, the target node of a negative example is quite likely to have an in-degree of one.
These death information indices make use of the fact that when we sample a positive example, we are going to decrease the in-degree of the target to zero, and when we sample a negative example, the in-degree of the target is likely to still be one (i.e. that the target was killed by somebody else at some point).
We try to account for this by adding some additional isolated nodes (corresponding to characters that did not kill anyone and did not die), but the index still performs best on the expanded network.
So although the baseline index and the other indices which use death information achieve good results, they do not do so by using any structural properties of the network but rather just by abusing the way our network is constructed.
The key takeaway here is that achieving an AUC score around $0.85$ is not that hard. The hardest part is achieving a noticeable increase in performance over the baseline classifier.
The results from the standard machine learning approaches are bad, since we only use the out degree as the basic feature from the original dataset to predict kills. 
When we augment it with different centrality measures and community identifiers, the performance is improved, mostly because of the fact that we do not only have one feature anymore and because these features are not equally weighted anymore. 
However, there should be some added value to the features due to what the features represent, which is explained for each index in Section~\ref{sec:methods}.

\begin{figure*}
    \centering
    \includegraphics[width=0.8\textwidth]{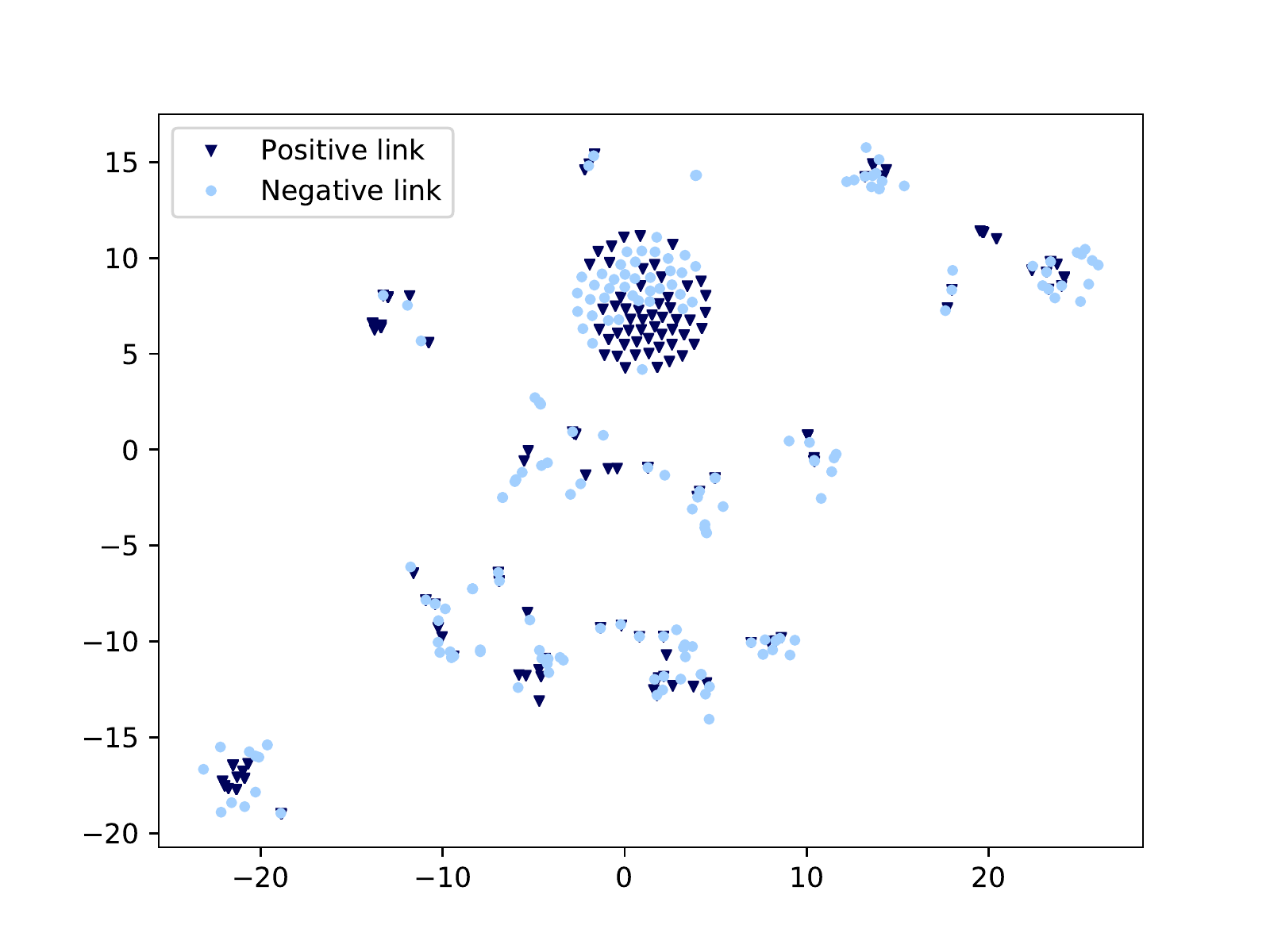}
    \caption{Node2vec embeddings for positive and negative links in the kills network, projected onto two dimensions using t-SNE.}
    \label{fig:node2vec-links-tsne}
\end{figure*}

Among the approaches using node2vec features, the approach using KNN achieves the best results.
Visualization of the link embeddings reveals why the approach performs well. 
Figure~\ref{fig:node2vec-links-tsne} shows embedded positive and negative links, projected onto a two-dimensional plane using t-distributed Stochastic Neighbor Embedding (t-SNE) ~\cite{maaten2008visualizing}.
We can observe that the links are often surrounded by links of the same class in their neighbourhood, allowing KNN to correctly predict many links.
The visualization also reveals one of the reasons the approaches using node2vec do not perform better: the links where at least one of the nodes does not have a ``proper'' embedding are embedded closely.
The circle-shaped cluster in the visualization represents the links where at least one of the nodes were assigned a generic embedding for unknown nodes (due to some character being present in the social network but not in the kills network). 
The cluster contains very mixed classes, rendering KNN less useful.
In additional experiments using node2vec features, we found that using a link embedding technique where node embeddings are averaged instead of concatenated together (i.e. ignoring direction of the link) results in a significant performance drop.
The modified embeddings in combination with KNN result in a mean AUC reduced by $0.068$, a mean precision reduced by $0.028$ and mean recall reduced by $0.127$.
This aligns well with intuition since, for example, a notorious killer is more likely to kill an innocent victim than vice versa.

\section{\label{sec:conclusion}Conclusion}

Throughout our work we have acknowledged that the Game of Thrones kills do not have a particularly detectable pattern, since all the kills appear between two nodes that have not yet killed anyone before in most cases. 
But since our network is small, our test samples are even smaller and that can give deceivingly high AUC scores for indices that would potentially fail on bigger networks and in different scenarios.
By using classic machine learning and link prediction techniques, we have found that, on this dataset, no index or feature works better than a simple baseline index (the alive index), which does not model some useful property, but rather abuses the way the network is formed.
Most of the techniques used to predict kills in Game of Thrones gave us a fairly good AUC score, but the predictions that our approaches get right do not have a high ``shock factor``.
For example, our classifiers might be able to predict that Jon Snow will kill a wight, but likely fails on less obvious kills.
For future work, our approaches could be tested on different TV shows, books or even movies to see how predictable the kills are. 
Our death information indices could potentially fail on bigger networks with a bit more diverse structure (e.g. having more bigger connected components) and that would give other more basic indices higher accuracy. 
We could also construct edges using some other information besides kills, e.g. who had a relationship with whom in some show or who scammed whom in a criminal series.

\subsection*{\label{sec:source}Source code}

The source code to reproduce results presented in this paper is available at \url{https://github.com/matejklemen/got-link-prediction}.

%
%

\bibliography{bibliography}

\begin{thebibliography}{22}
\providecommand{\natexlab}[1]{#1}
\providecommand{\url}[1]{\texttt{#1}}
\expandafter\ifx\csname urlstyle\endcsname\relax
  \providecommand{\doi}[1]{doi: #1}\else
  \providecommand{\doi}{doi: \begingroup \urlstyle{rm}\Url}\fi

\bibitem[Liben-Nowell and Kleinberg(2007)]{LNK07}
David Liben-Nowell and Jon Kleinberg.
\newblock The link-prediction problem for social networks.
\newblock \emph{Journal of the American society for Information Science and
  Technology}, 58\penalty0 (7):\penalty0 1019--1031, 2007.

\bibitem[Adamic and Adar(2003)]{AA03}
Lada~A Adamic and Eytan Adar.
\newblock Friends and neighbors on the web.
\newblock \emph{Social Networks}, 25\penalty0 (3):\penalty0 211--230, 2003.

\bibitem[Barab{\'a}si and Albert(1999)]{BA99}
Albert-L{\'a}szl{\'o} Barab{\'a}si and R{\'e}ka Albert.
\newblock Emergence of {S}caling in {R}andom {N}etworks.
\newblock \emph{Science}, 286\penalty0 (5439):\penalty0 509--512, 1999.

\bibitem[Albert and Barab{\'a}si(2002)]{BA02}
R{\'e}ka Albert and Albert-L{\'a}szl{\'o} Barab{\'a}si.
\newblock Statistical mechanics of complex networks.
\newblock \emph{Reviews of modern physics}, 74\penalty0 (1):\penalty0 47, 2002.

\bibitem[Kossinets(2006)]{K06}
Gueorgi Kossinets.
\newblock Effects of missing data in social networks.
\newblock \emph{Social networks}, 28\penalty0 (3):\penalty0 247--268, 2006.

\bibitem[Guimer{\`a} and Sales-Pardo(2009)]{GS09}
Roger Guimer{\`a} and Marta Sales-Pardo.
\newblock Missing and spurious interactions and the reconstruction of complex
  networks.
\newblock \emph{Proceedings of the National Academy of Sciences}, 106\penalty0
  (52):\penalty0 22073--22078, 2009.

\bibitem[L{\"u} and Zhou(2011)]{LZ11}
Linyuan L{\"u} and Tao Zhou.
\newblock Link prediction in complex networks: A survey.
\newblock \emph{Physica A: statistical mechanics and its applications},
  390\penalty0 (6):\penalty0 1150--1170, 2011.

\bibitem[Girvan and Newman(2002)]{GN02}
Michelle Girvan and Mark~EJ Newman.
\newblock Community structure in social and biological networks.
\newblock \emph{Proceedings of the National Academy of Sciences}, 99\penalty0
  (12):\penalty0 7821--7826, 2002.

\bibitem[Newman(2001)]{N01}
Mark~EJ Newman.
\newblock Clustering and preferential attachment in growing networks.
\newblock \emph{Physical review E}, 64\penalty0 (2):\penalty0 025102, 2001.

\bibitem[McGill and Salton(1983)]{MS83}
M~McGill and G~Salton.
\newblock Introduction to modern information retrieval. 1983.
\newblock \emph{McGraw-Hil, New York}, 1983.

\bibitem[Jaccard(1901)]{J01}
Paul Jaccard.
\newblock {\'E}tude comparative de la distribution florale dans une portion des
  alpes et des jura.
\newblock \emph{Bull Soc Vaudoise Sci Nat}, 37:\penalty0 547--579, 1901.

\bibitem[Liu and Albergante(2017)]{LA17}
Dianbo Liu and Luca Albergante.
\newblock Balance of thrones: a network study on "{G}ame of {T}hrones" that
  unveils predictable popularity of the story.
\newblock \emph{arXiv preprint arXiv:1707.05213}, 2017.

\bibitem[Beveridge and Shan(2016)]{BS16}
Andrew Beveridge and Jie Shan.
\newblock Network of thrones.
\newblock \emph{Math Horizons}, 23\penalty0 (4):\penalty0 18--22, 2016.

\bibitem[Angraal et~al.(2018)Angraal, Bhatnagar, Verma, Shergill, Gupta, and
  Khera]{ABSSAR2018}
Suveen Angraal, Ambika Bhatnagar, Suraj Verma, Sukhman Shergill, Aakriti Gupta,
  and Rohan Khera.
\newblock Risk {F}actors {A}ssociated with {M}ortality in {G}ame of {T}hrones:
  {A} {L}ongitudinal {C}ohort {S}tudy.
\newblock \emph{arXiv preprint arXiv:1802.04161}, 2018.

\bibitem[Cox(1972)]{C72}
David~R Cox.
\newblock Regression models and life-tables.
\newblock \emph{Journal of the Royal Statistical Society: Series B
  (Methodological)}, 34\penalty0 (2):\penalty0 187--202, 1972.

\bibitem[Watts and Strogatz(1998)]{WS98}
Duncan~J. Watts and Steven~H. Strogatz.
\newblock {Collective dynamics of 'small-world' networks}.
\newblock \emph{Nature}, 393\penalty0 (6684):\penalty0 440--442, June 1998.

\bibitem[Traag et~al.(2019)Traag, Waltman, and van Eck]{TWVE18}
Vincent~Antonio Traag, Ludo Waltman, and Nees~Jan van Eck.
\newblock From {Louvain} to {Leiden}: guaranteeing well-connected communities.
\newblock \emph{Scientific Reports}, 9\penalty0 (1):\penalty0 5233, 2019.
\newblock \doi{10.1038/s41598-019-41695-z}.

\bibitem[Brin and Page(1998)]{BP98}
Sergey Brin and Lawrence Page.
\newblock The {A}natomy of a {L}arge-{S}cale {H}ypertextual {W}eb {S}earch
  {E}ngine.
\newblock In \emph{Seventh International World-Wide Web Conference}, 1998.

\bibitem[Hastie et~al.(2009)Hastie, Tibshirani, and Friedman]{HTF09}
Trevor Hastie, Robert Tibshirani, and Jerome Friedman.
\newblock \emph{{The Elements of Statistical Learning: Data Mining, Inference,
  and Prediction}}.
\newblock Springer, 2nd edition, 2009.

\bibitem[Freeman(1977)]{F77}
Linton~C Freeman.
\newblock A set of measures of centrality based on betweenness.
\newblock \emph{Sociometry}, pages 35--41, 1977.

\bibitem[Grover and Leskovec(2016)]{grover2016node2vec}
Aditya Grover and Jure Leskovec.
\newblock node2vec: Scalable feature learning for networks.
\newblock In \emph{Proceedings of the 22nd ACM SIGKDD international conference
  on Knowledge discovery and data mining}, pages 855--864, 2016.

\bibitem[Maaten and Hinton(2008)]{maaten2008visualizing}
Laurens van~der Maaten and Geoffrey Hinton.
\newblock Visualizing data using t-sne.
\newblock \emph{Journal of machine learning research}, 9\penalty0
  (Nov):\penalty0 2579--2605, 2008.

\end{thebibliography}

\end{document}